\documentclass[prd,twocolumn,a4paper,superscriptaddress,floatfix]{revtex4-1}
\usepackage{graphicx}
\begin{document}

\newcommand{\be}{\begin{equation}}
\newcommand{\ee}{\end{equation}}
\newcommand{\bq}{\begin{eqnarray}}
\newcommand{\eq}{\end{eqnarray}}
\newcommand{\bsq}{\begin{subequations}}
\newcommand{\esq}{\end{subequations}}
\newcommand{\bc}{\begin{center}}
\newcommand{\ec}{\end{center}}

\title{Evolution of Semilocal String Networks: I. Large-scale Properties}
\author{A. Ach\'ucarro}
\email{achucar@lorentz.leidenuniv.nl}
\affiliation{Institute Lorentz of Theoretical Physics, University of Leiden, 2333CA Leiden, The Netherlands}
\affiliation{Department of Theoretical Physics, University of the Basque Country UPV-EHU, 48040 Bilbao, Spain}
\author{A. Avgoustidis}
\email{tavgoust@gmail.com}
\affiliation{School of Physics and Astronomy, University of Nottingham, University Park, Nottingham NG7 2RD, England}
\author{A. M. M. Leite}
\email{up080322016@alunos.fc.up.pt}
\affiliation{Centro de Astrof\'{\i}sica, Universidade do Porto, Rua das Estrelas, 4150-762 Porto, Portugal}
\affiliation{\'Ecole Polytechnique, 91128 Palaiseau Cedex, France}
\author{A. Lopez-Eiguren}
\email{asier.lopez@ehu.es}
\affiliation{Department of Theoretical Physics, University of the Basque Country UPV-EHU, 48040 Bilbao, Spain}
\author{C. J. A. P. Martins}
\email{Carlos.Martins@astro.up.pt}
\affiliation{Centro de Astrof\'{\i}sica, Universidade do Porto, Rua das Estrelas, 4150-762 Porto, Portugal}
\author{A. S. Nunes}
\email{c0808014@alunos.fc.up.pt}
\affiliation{Centro de Astrof\'{\i}sica, Universidade do Porto, Rua das Estrelas, 4150-762 Porto, Portugal}
\affiliation{Faculdade de Ci\^encias da Universidade de Lisboa, Campo Grande, 1749-016 Lisboa, Portugal}
\author{J. Urrestilla}
\email{jon.urrestilla@ehu.es}
\affiliation{Department of Theoretical Physics, University of the Basque Country UPV-EHU, 48040 Bilbao, Spain}

\date{4 February 2014}
\begin{abstract}
We report on a detailed numerical study of the evolution of semilocal string networks, based on the largest and most accurate field theory simulations of these objects to date. We focus on the large-scale network properties, confirming earlier indications (coming from smaller simulations) that linear scaling is the attractor solution for the entire parameter space of initial conditions that we are able to probe. We also provide a brief comparison of our numerical results with the predictions of a previously developed one-scale model for the overall evolution of these networks. Two subsequent papers will discuss in more detail the analytic modeling of the semilocal segment populations as well as optimized numerical diagnostics.
\end{abstract}
\maketitle

\section{Introduction}
\label{intro}

The formation of networks of cosmic strings \cite{Kibble80,HindKib,VilShell,CopKib} is a generic prediction 
in a wide range of high-energy physics models of the early universe \cite{VilShell,Jeannerot,DvalTye}. 
Examples include line-like topological defects in field theories breaking $U(1)$ symmetry \cite{NielOl}, 
coherent macroscopic states of fundamental superstrings (F-strings), 
and D-branes extended in one macroscopic direction (D-strings). The latter two examples, collectively 
referred to as cosmic superstrings \cite{PolchIntro}, are generically predicted in string theoretic inflationary models involving 
spacetime-wrapping D-branes \cite{BMNQRZ,MajumDavis,DvalTye,PolchStab}. 
These `Brane Inflation' models often have an effective Supergravity description of the hybrid inflation type, ending 
in a phase transition that produces topological defects, so F- and D-strings can also be modeled as string 
defects in a field theory approximation. 

A key property of cosmic superstrings is that they interact non-trivially, joining together in Y-shaped 
junctions to form heavier bound FD-states \cite{Schwarz,Witten95,PolchIntro}, and in this respect they 
are similar to non-Abelian strings \cite{VilShell}. There is, however, an important distinction between string 
defects in ordinary 4D field theories and their higher-dimensional superstring cousins: field theoretic strings 
are known to interact with probabilities of order unity \cite{Shellard87,Matzner} (although at
ultra-relativistic speeds the strings can appear to pass through each other due to multiple intercommutations
\cite{Achucarro:2006es,Achucarro:2010ub,Verbiest:2011kv}), while cosmic superstrings 
have intercommutation probabilities which can be much smaller than unity \cite{JoStoTye,PolchProb}. This 
has an important effect on the cosmological evolution of superstring networks leading to higher number 
densities than for ordinary field theory cosmic strings \cite{JoStoTye,Sak04,Avg05}. 

Understanding the evolution of string networks is crucial for predicting their number densities 
at late times, which in turn determine their potentially observable effects. Since these observational 
signals depend on parameters of the underlining theory (notably through their sensitivity to the string 
tensions and intercommutation probabilities), the observational search for cosmic strings provides a 
powerful tool for probing and/or constraining high-energy physics theories of the early 
universe \cite{LorMartRing,CopPogVach, HindmRev}. However, the quantitatively accurate modeling 
of string network evolution is a difficult problem, requiring the combination of a range of techniques 
(both numerical and analytical), and interpolating between physics at very different energy scales. 

For the simplest type of Abelian cosmic strings (e.g. the Nielsen-Olesen solutions of the Abelian 
Higgs model \cite{NielOl}) which interact simply by exchange of partners, it has long been proposed 
\cite{Kibble85} through analytic modeling that the network should reach a self-similar scaling regime, 
characterised by a single length-scale (the correlation length) which asymptotes to a constant fraction 
of the horizon. This has been confirmed by high-resolution numerical simulations of both 
Nambu-Goto \cite{BenBouch,AllShell,Fractal} and field theory \cite{VincAntHind,YamYokKaw,Moore,Bevis} 
models, which are in remarkable agreement despite the very different techniques employed in each 
case. Even in this simplest type of strings, there still remains significant numerical uncertainty regarding the 
relevant importance of decay mechanisms (gravitational radiation vs decay to particles \cite{Moore,Stuckey}) and the 
average size of loops in the network \cite{Fractal,Ringeval,Vanchurin,BlancoPillado}, but consensus has long been reached regarding the large 
scale properties of long strings and their quantitative dependence on the intercommutation probability 
and the rate of cosmic expansion.        

The situation is less clear for non-Abelian strings which do not simply exchange partners but interact 
in a more complex fashion, forming Y-type junction configurations. This was originally though to lead 
to network ``frustration" implying a cosmologically catastrophic domination of strings over ordinary matter 
at late times \cite{SperPen}. Subsequent work, however, has indicated that this is not necessarily the 
case \cite{McGraw}, and whether the network reaches scaling or gets frustrated depends on the relation 
among the various intercommutation probabilities of strings carrying different charges \cite{NAVOS}. 
In particular, for networks resembling the properties of cosmic superstrings, all recent studies (see for 
example \cite{TWW,CopSaf,Saffin,HindSaf,NAVOS,UrrVil,SakStoic,AvgCop,PACPS}), covering both analytic 
and field theory modeling, have found scaling solutions with the relative abundance of light F-strings 
dominating over the heavier D-strings and FD bound states. Thus, it is now believed that cosmic superstring 
networks do reach late time scaling with light strings being more abundant, even though to date it has not 
been possible to construct both analytic and field theory models of the 'same' network so as to 
quantitatively compare their abundance predictions.       

There is an outstanding case of cosmic string networks whose cosmological evolution has not been 
systematically studied: semilocal strings. These are string solutions in theories with both local and 
global symmetries, the standard semilocal model \cite{VachAchuc,Achucarro:1999it} being a minimal extension 
of the Abelian Higgs model by a global SU(2) symmetry. This model, which has an SU(2) doublet of two equally charged Higgs fields 
under a single U(1) gauge field, admits stable string solutions even though the vacuum manifold is 
simply connected. This non-topological nature of semilocal strings endows them with very different 
properties than their topological counterparts. In particular, they appear as finite open segments 
whose ends have long-range interactions akin to global monopoles \cite{Hindmarsh93}.
Note that such strings are also well-motivated from the theoretical point of view, arising in supersymmetric grand unified theories of 
inflation \cite{Urrestilla:2004eh} and the corresponding D3/D7 brane inflation models \cite{D3D7semiloc}. 
These are a natural extension of usual inflationary models, in which the only extra ingredient is the 
doubling of a hypermultiplet.     

A first study of the cosmological evolution of semilocal strings was presented in~\cite{Achucarro:1998ux} and \cite{Achucarro:2007sp}. 
The dynamics of these networks is very different than for both Abelian and non-Abelian topological strings, 
be it global or local. In particular, the long range forces between the monopoles mean that the segments can 
shrink and annihilate or grow by joining with other segments. In a recent paper \cite{Nunes} we initiated the 
analytical study of such networks by modeling them as local strings ending on global monopoles, and 
attempted a preliminary comparison with numerical simulations. Here, we present the first detailed 
numerical study of semilocal string networks. In this paper (Paper I), which is the first in a series of three, 
we will discuss in detail the large-scale properties of simulated semilocal networks, covering couplings 
in the range $0.01\le \beta \le 0.09$, and damping terms corresponding to expanding universes dominated by
radiation and matter.

Our goal is to demonstrate scaling behavior for semilocal networks. We note, however, that the notion of scaling has to be interpreted carefully in this context. When describing these networks we may simply be interested in the evolution of the overall energy density contained in the semilocal string configurations, or we may be interested in the detailed distribution of semilocal string segments. The former (which will be the focus of this paper) is the simplest in the sense that it can, to a good approximation, be described by a single lengthscale, while the latter is somewhat more complex. We will explore this distinction further in the subsequent papers of this series, but for the moment we emphasize that overall scaling of the network's energy density is necessary but not sufficient to ensure scaling of the segment distribution.

In Paper II, we will present the results of detailed comparison of our simulations with the analytical models 
of Ref.~\cite{Nunes}: starting from an initial configuration of a semilocal network we group all string segments into length bins and evolve the segments in each bin both using our field theory simulation and our analytic models. We then compare the number density in each bin between the two approaches.         
An important source of uncertainly in this comparison is related to our lack of knowledge of the transverse 
string velocity in simulations of semilocal strings. In Paper III we will present a novel method for measuring velocities from semilocal string simulations.   

\section{Semilocal Strings}
\label{semi}

Semilocal strings \cite{Achucarro:1999it,Hindmarsh:1991jq,VachAchuc} were introduced as a minimal extension of the Abelian Higgs model with two complex scalar fields---instead of just one---that make an $SU(2)$ doublet. This leads to $U(1)$ flux-tube solutions even though the vacuum manifold is simply connected. The strings of this extended model have some similarities with ordinary local $U(1)$ strings, but they are not purely topological and will therefore have different properties. For example, since they are not topological, they need not be closed or infinite and can have ends. These ends are effectively global monopoles with long-range interactions \cite{Hindmarsh93} that can make the segments grow or shrink. The monopoles at the ends of the strings have some exotic properties by themselves~\cite{Achucarro:2000td}.

The relevant action for the simplest semilocal string model, the one
we will use in the numerical simulations of section~\ref{comp}, reads
\be
S=\int\!\! d^4x\!\left[\left[(\partial_\mu-i
    A_\mu)\Phi\right]^2-\frac{1}{4}F^2-\frac{\beta}{2}(\Phi^+\Phi-1)^2\right]
\label{SLaction}
\ee
where $\Phi=(\phi,\psi)$, $F^2 = F_{\mu \nu} F^{\mu \nu}$ and
$F_{\mu\nu} = (\partial_\mu A_\nu - \partial_\nu A_\mu)$ is the gauge
field strength. It can be easily seen that setting one of the two
scalar fields to zero, we recover the Abelian Higgs model. We can
therefore build from the analytical models applied to usual cosmic
strings to tackle this new problem.

The symmetry breaking pattern that leads to the formation of
strings in this model is $SU(2)_{\rm global} \times U(1)_{\rm local} \to
U(1)_{\rm global}$ so this model can be thought of as a particular limit
of the Glashow-Weinberg-Salam electroweak model in which the
SU(2) symmetry is global, i.e.  the Weinberg angle is $\cos\theta_W =
0$ and there are no SU(2) gauge fields.  The vacuum manifold is the
three sphere, so one would not expect strings to form if the dynamics
is dominated by the potential energy. On the other hand, the magnetic
field is massive and magnetic flux is conserved, which would suggest
the existence of magnetic flux tubes when the magnetic mass is
large. This is the regime in which strings form and are stable.

The stability of the strings is not trivial, and it will depend on the
value of the parameter $\beta=m^2_{\rm scalar}/m^2_{\rm gauge}$: for
$\beta<1$ the string is stable, for $\beta>1$ it is unstable, and for
$\beta=1$ it is neutrally stable \cite{Hindmarsh:1991jq,VachAchuc}. As
we will see in section~\ref{comp}, only low values of $\beta$ will be
of interest for the comparison, because otherwise the string network
is either unstable or disappears very fast~\cite{Achucarro:1997cx,Achucarro:1998ux,Achucarro:2007sp}.

After a cosmological phase transition in such a model, it is expected
that segments of semilocal strings will form. The cosmological
evolution of a semilocal segment network will be quite different from
the evolution of ordinary Abelian-Higgs
strings~\cite{Benson:1993at,Achucarro:1999it}. The fact that semilocal
strings have a different cosmological evolution is interesting because
CMB predictions can be different~\cite{Urrestilla:2007sf} and can be relevant to inflationary model
building~\cite{Urrestilla:2004eh}.  Semilocal strings also have interesting gravitational
properties~\cite{Hartmann:2009tc,Hartmann:2012ad}.

The network evolution depends on the interplay between string dynamics
and monopole dynamics.  When a string segment ends, it must end in a
cloud of gradient energy. Those string ends behave like global
monopoles providing an interaction between strings that is
independent of distance. Therefore, depending on the interplay between
string dynamics and monopole dynamics, the segments can contract and
eventually disappear, or they can grow to join a nearby segment and
form a very long string, and also the two ends of a segment can
join to form a closed loop~\cite{Hindmarsh93,Achucarro:1998ux,Achucarro:2007sp}.

We thus see that, at least to a first approximation, we can envisage these networks as being made of local strings attached to global monopoles, and, as such, previously developed analytic modeling techniques~\cite{VOS96,MartAch} should be applicable. This being said, it is also clear that these networks possess additional dynamical properties, beyond those of standard hybrid networks \cite{VOS96,MartAch,Hybrid}.

Specifically, the evolution of the string network will depend both on the string tension and on the dynamics of the gradient energy: the latter may be thought of as providing a long-range interaction between the strings. (Note that the force between global monopoles is independent of distance.) In Ref.~\cite{Nunes} we presented analytical models for the cosmological evolution of semilocal networks, taking into account these long-range interactions through the addition of phenomenological terms in hybrid (local strings + global monopole) networks. We provide a quick summary of this analytical approach in the next section, before moving to the presentation of our numerical study on Section~\ref{numerics}.     

\section{Semilocal Network Evolution Modeling}
\label{networks}

We now discuss an analytic model for the evolution of semilocal string networks, which will be subsequently compared to our numerical simulations. This is mostly a summary of \cite{Nunes}, where the model was first presented; we refer the reader to that work for additional details. 

Our analysis focuses on the behavior of the network as a whole, starting from the premise that it can be treated as a network of local strings attached to global monopoles. Therefore previously developed models for each of these cases can be applied, with suitable changes, to this case. Our model for the evolution of these networks is based on explicitly modeling the dynamics and interactions of the monopoles. This is justified since (as has been shown in previous work \cite{Achucarro:2007sp}) it is indeed the monopoles that control the evolution of the network.

A complementary approach (also developed in \cite{Nunes}) models the evolution of individual semilocal segments, discussing under what conditions these segments can grow---a process that has been clearly identified in numerical simulations---or shrink. We will not discuss this here since a detailed study of this approach, including comparisons with the numerical simulations discussed in this paper, will be the focus of Paper II.

Analytic modeling of defect networks generally starts from the microscopic equations of motion for the defects and uses statistical averaging procedures leading to a macroscopic energy evolution equation (which can be traded for an equation for the network's characteristic lengthscale) and an equation for the RMS network velocity. These equations will necessarily be coupled, and together they describe the network at large-scales in a `thermodynamical' sense. Suitable defect interactions are then added to these equations in a phenomenological way. This procedure was originally followed in the case of cosmic strings, where it leads to the so-called velocity-dependent one-scale (VOS) model \cite{VOS96,VOS02}, which has been well-tested against simulations and is used for predicting CMB signals of string networks \cite{PlanckDefects}.   
    
Similar techniques can be used in the case of monopoles.  The idea is to obtain an 
evolution equation for the monopole density (neglecting interactions) and then re-express it
in terms of a characteristic lengthscale, $L$ (which in this case should be thought of as the average
inter-monopole distance). The effects of monopole forces and friction are then included in this equation
(as well as in the relevant velocity equation) by adding extra phenomenological terms. It has been shown
in \cite{MartAch,Hybrid} that the evolution equation for the characteristic monopole lengthscale has the form
\be
3\frac{dL}{dt}=3HL+v^2\frac{L}{\ell_d}+c_\star v\,,
\ee
where $c_\star$ is a free parameter (to be calibrated by simulations) quantifying energy loss, and where 
we have defined a damping lengthscale, $l_d$ that includes the effects of expansion (due to the Hubble parameter $H$) 
and of friction due to particle scattering (with a generic lengthscale $l_f$)
\be
\frac{1}{l_{d}}=H+\frac{1}{l_{f}}\,.
\ee
The evolution equation for the RMS velocity $v$ of the monopoles is
\be\label{vosv}
\frac{dv}{dt}=(1-v^2)\left[\frac{k_m}{L}\left(\frac{L}{d_H}\right)^{3/2}+\frac{k_s}{L}\frac{\eta_s^2}{\eta_m^2}-\frac{v}{\ell_d}\right]\,,
\ee
where the first term in square brackets is the force due to the monopoles
\be
f_m=\frac{k_m}{L}\left(\frac{L}{d_H}\right)^{3/2}
\ee
and the second describes the force due to the strings
\be
f_s=\frac{k_s}{L}\frac{\eta_s^2}{\eta_m^2}\,.
\ee
For an expansion rate of the generic form
\be
a(t)\propto t^{\lambda}\,
\ee
the Hubble parameter and horizon distance are respectively 
\be
H=\frac{\lambda}{t}\,,\quad d_{H}=\frac{t}{1-\lambda}\,.
\ee
The constants $k_m$ and $k_s$ parametrize the monopole and string forces, and $\eta_s$, $\eta_m$ 
are the relevant symmetry breaking scales. Since in what follows we are mostly interested in late-time scaling
solutions we will (unless  otherwise stated) neglect the effect of friction due to particle scattering, which is
only relevant in the early stages of the network's evolution.

Note that the fact that the string and monopole symmetry breaking scales appear in Eq. (\ref{vosv}) is a consequence of the fact that these equations of motion are obtained by modeling semilocal strings as local strings attached to global monopoles (as previously mentioned), and appropriately adapting the equations of motion for both. Physically one knows that it is the monopoles that dominate the semilocal string dynamics, and this can be modeled by assuming that $\eta_s\ll\eta_m$. Similarly, the horizon enters in the monopole force term in Eq. (\ref{vosv}) due to a number counting argument: this force depends on the number of monopoles (and antimonopoles) inside the horizon; for a detailed discussion see \cite{MartAch} and references therein.

We shall mostly consider standard expansion rates, corresponding to the parameter range $0<\lambda <1$, and in particular $\lambda=1/2$ in the radiation-dominated era and $\lambda=2/3$ in the matter-dominated era. This is justified since observational constraints \cite{Urrestilla:2007sf,PlanckDefects} show that semilocal string networks cannot be the dominant component of the universe's energy budget, but can only contribute a small fraction to it.

In the semilocal case the ratio of the forces due to strings and monopoles is
\be
\frac{f_s}{f_m}=\frac{k_s}{k_m}\left(\frac{\eta_s}{\eta_m}\right)^2\left(\frac{d_H}{L}\right)^{3/2}
\ee
and since $\eta_s\ll\eta_m$ the string force is always subdominant. This is in agreement with theoretical expectations and numerical simulations. Note that this is a distinguishing characteristic of these networks: for local strings attached to local monopoles the force due to the strings always dominates the dynamics, while for global strings attached to global monopoles the string force is subdominant at string formation but becomes dominant later in the network's evolution \cite{Hybrid}. 

One interesting consequence of the fact that the monopoles always dominate the dynamics is that the only attractor solution of these evolution equations in an expanding universe (with $a\propto t^\lambda$) is linear scaling. Indeed, if one looks for generic solutions of the equations of motion for $L$ and $v$, allowing for arbitrary power laws of time in eitheir case, one will find (after a relatively long but otherwise straightforward calculation) that the only consistent asymptotic solution is
\be
L=\gamma t\,,\quad v=v_0\,,\label{linscal}
\ee
as in the case of plain global monopoles, and indeed the analysis in \cite{MartAch} is, to a large extent, applicable here.

There are two possible branches of the scaling solution. First, there is an ultra-relativistic one with
\be
\gamma=\frac{c_\star}{3-4\lambda}\,,\quad v_0=1\,,
\ee
which only exists for slow expansion rates ($\lambda<3/4$) but is in principle allowed both on the radiation and matter eras. Second, a normal solution exists for any expansion rate, with scaling parameters
\be
\gamma = \frac{c_\star v_0}{3-\lambda(3+v_{0}^{2})} \, \label{nscaling1}
\ee
\be
\lambda v_0=k_m(1-\lambda)^{3/2}\gamma^{1/2}\,, \label{nscaling2}
\ee
and a constraint on the velocities
\be
v_0^2<3(\frac{1}{\lambda}-1)\,.
\ee
This constraint is trivial for $\lambda<3/4$ (that is, $v_0\to1$ is allowed), but restrictive for faster expansion rates. On the other hand, velocities will generically be significant: having $v_0\to0$ requires $\lambda\to1$. 

For comparison we also consider the case of Minkowski space (corresponding to $\lambda=0$ and $H=0$) but with a friction lengthscale proportional to the correlation length (say, for simplicity, $\ell_f\sim L$). This should be an adequate description of some of the numerical simulations of semilocal strings done so far \cite{Achucarro:2007sp}. In this case, linear scaling is still the attractor solution but the scaling parameters now obey
\be
3\gamma =v_0^2+cv_0\,,\quad v_0=k_m\gamma^{3/2}\,.
\ee

In the opposite limit of fast expansion rate ($\lambda\ge1$, or in other words inflation) the linear scaling solution of Eq.~(\ref{linscal}) no longer exists. In this case the network is conformally stretched and gradually frozen, and the characteristic lengthscale and velocity evolve as
\be
L\propto a\,,\quad v\propto \frac{1}{HL}\,.
\ee
These conformal stretching solutions are ubiquitous in the defects literature.

In the following section we will test these scaling solutions using state-of-the-art numerical simulations.

\section{Numerical simulations}
\label{numerics}

We simulated numerically the semilocal model introduced in section~\ref{semi} so as to provide us with data to be used for comparison with the analytic models mentioned above. The parameter space we want to explore is rather large, so we carefully chose the cases to study, and tried to maximize the information we could obtain from our simulation given the computer resources available to us.

We discretized the action given in equation~(\ref{SLaction}) by standard techniques (using lattice-link variables and a staggered-leapfrog method) and evolved the discretized action in 1024$^3$ lattices with periodic boundary conditions, similar to \cite{Achucarro:2007sp}. One very important approximation we use in our discretization and subsequent evolution of the equations of motion is the use of the so-called \emph{fat-string} algorithm \cite{Press:1989yh}. We adopt this approach  since otherwise it would be computationally very expensive to perform the simulations, and because it has proved to work fairly well in previous works; in particular, it has been shown that it works fine for obtaining information on large-scale properties, which is our aim in the present work. (A related discussion for the case of cosmic strings can be found in \cite{Hiramatsu}.)

As in many field theoretic simulations of defect dynamics, the initial conditions are an unknown. It would be very hard to simulate exactly the phase transition leading to the formation of the defects, and in many cases it would not be clear which model to adopt as the underlying phase transition model. However, this is not the goal; instead, our aim is to study the asymptotic (long-term) behavior of these networks, and in particular whether (and under what conditions) the expected scaling solution is reached. The art of performing the simulations therefore lies in obtaining some initial conditions which may approach this putative scaling solution as fast as possible. Bear in mind that the periodic boundary conditions force us to have a stringent upper bound on the time that the system can be evolved before it feels the effects of the boundaries. The simulations can only be believed up to the half light-crossing time, \emph{i.e.}, if we sent a light ray in opposite directions in the box, the simulation would be accurate up to when the two rays meet again.

The initial condition chosen for these simulation is the following: the gauge field, gauge field velocities and scalar field velocities are set to zero. This choice already ensures that Gauss's law is satisfied in the discretized equations and will be satisfied during the dynamical evolution of the system. The scalar fields are chosen to lie in the vacuum manifold, but have randomly chosen orientations. After a transient time with an ad-hoc damping term for the system to lose energy, the system relaxes into the scaling regime. 

Once the system reaches scaling, quantities of interest can be measured. Semilocal strings are not topological entities; therefore, we cannot use topology to detect semilocal strings. For example, in the usual Abelian-Higgs strings, one can use the windings of the string to pinpoint where the core of the string is. However, we cannot use the windings in the semilocal strings since the winding is not topologically protected. As mentioned earlier, semilocal strings can be though of as concentrations of magnetic energy, and that is the strategy we follow, inherited from previous works \cite{Urrestilla:2001dd,Pickles:2002ym,Achucarro:2007sp}: we first calculate the maximum of the magnetic field strength, and the radius,  of a straight and infinite Abelian Higgs string for a given $\beta$. We use those values for the simulated semilocal string network: if the magnetic field strength of a simulated semilocal model measured at a point of the  box exceeds the $25\%$ of the maximum of the corresponding Abelian-Higgs string, we consider that point to be part of a semilocal string segment.  The output of our simulation is thus an array of points from the simulated box which have a considerable concentration of magnetic field strength.

\begin{figure*}
\begin{center}
\includegraphics[width=15cm]{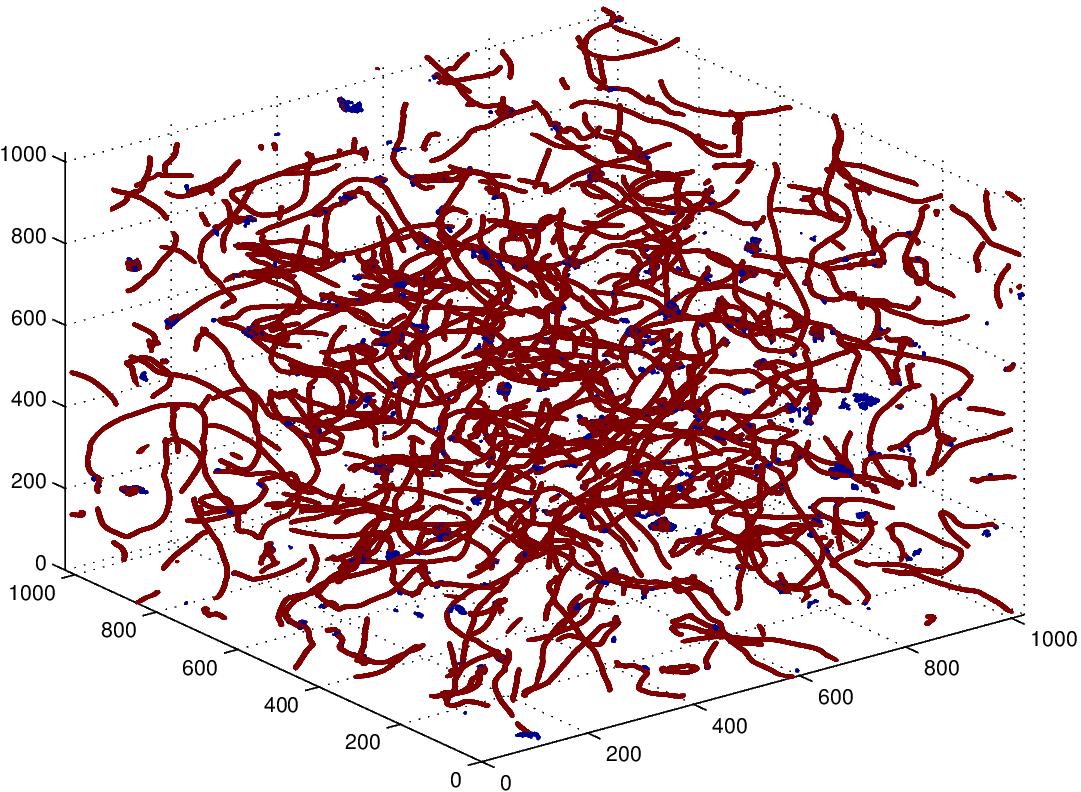}
\includegraphics[width=15cm]{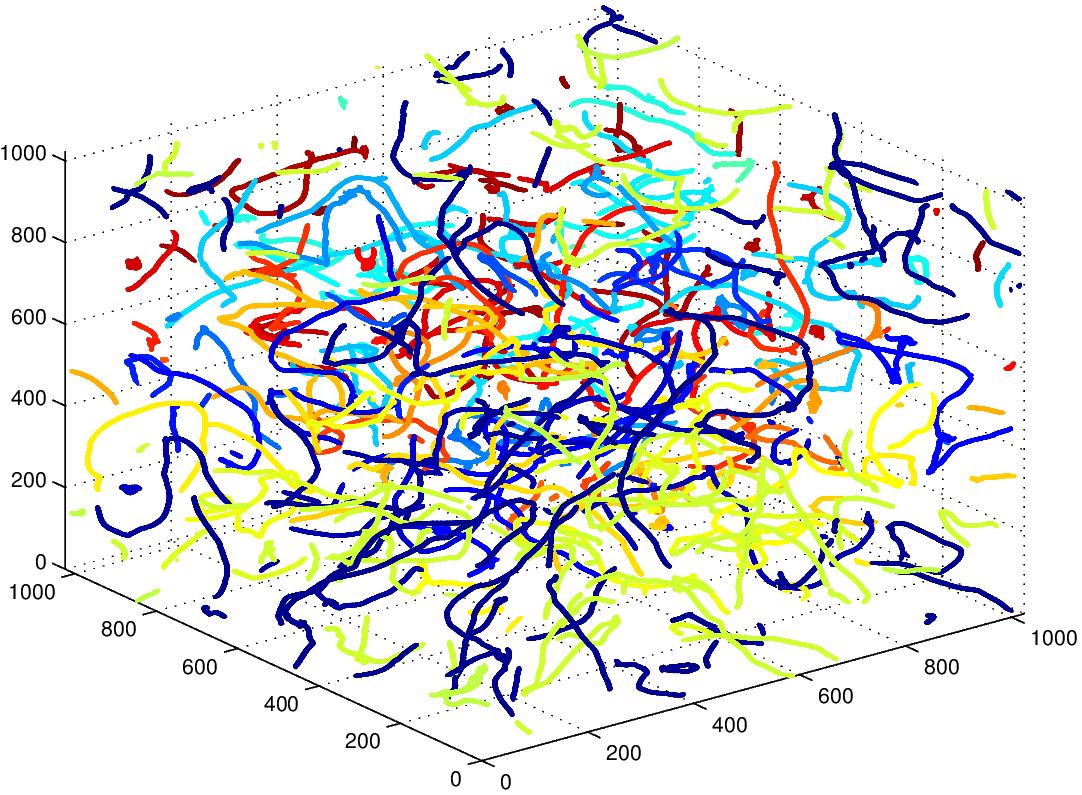}
\end{center}
\caption{Semilocal string network, in matter domination with $\beta=0.04$. The top figure shows two types of structures: on the one hand we have tube-like structures (proper strings) and on the other short blobs. These blobs we disregard in our analysis. The bottom figure shows the network without blobs, and also each segment has been identified and plotted with a different color. As there number of segments is large, the colours are unfortunately used for more than one string segment. Note also that the blob removal procedure does sometimes fail to identify some sphere-like structures, since their volume is large.}
\label{blobs}
\end{figure*}

One typical simulation snapshot is shown in Fig.~\ref{blobs}. It is very clear that semilocal strings have ends (as opposed to Abelian-Higgs strings which are either infinite or form loops). We then group together the points that have been output by the simulations  into segments. These segments are mostly tube-like, but some are sphere-like instead of tube-like,  i.e., they are {\it blobs} of energy. These can be formed, for example, after a segment has collapsed into itself. We do not wish to count these blobs as part of our network, and we introduce a lower cut-off: those segments that are not longer than a given factor ($\alpha$) times the typical radius of a string are considered to be blobs, and are discarded. Different choices of $\alpha$ have been considered as explained later. Fig.~\ref{blobs} shows the output of a typical simulation where we have differentiated between structures that we consider {\it blobs} and {\it proper} semilocal string segments. We also show in that figure the network of segments with each segment plotted  with a different colour.

It is now possible to obtain the necessary quantities for our comparison: the total string length (that is, the sum of all the segment lengths), the number of monopoles and the segment length distribution. The procedure we have described so far only gives the {\it volume} of string points, so in order to obtain string lengths, we divide the number of string points by the typical string width for each $\beta$. The number of monopoles is obtained by multiplying the number of segments by two, as each segment has a monopole and an antimonopoles at its ends (more on this point below). In what follows we do not directly compare the velocities in the model and simulations, since reliable numerical measurements of these velocities are highly non-trivial and require the development of additional numerical algorithms, which we will address in Paper III. For analogous issues in the more standard case of Abelian-Higgs string networks, see \cite{Moore}; for the case of domain wall simulations with the Press-Ryden-Spergel algorithm see \cite{WallSim}.

Given a box size (in our case 1024$^3$) one would want to have as big a dynamical range as possible, with as much accuracy as possible. There is clearly  tension between these two aims: on the one hand we would want a big lattice spacing ($\delta x$) to increase the dynamical range and on the other a small one to increase accuracy in the discretization. We have performed two sets of simulations trying to accommodate both needs: one set of simulations have $\delta x=0.5$  and the other $\delta x=1.0$. The first set provides a more accurate discretization of the equations, but pays the price of having a shorter dynamical range. The second has a larger dynamical range, but may lack in accuracy and there might be discretization effects creeping into the simulation. As will be shown below results obtained by the two approaches are clearly compatible, and we believe that they are accurate enough for the purposes of this paper. 

It is well known~\cite{Achucarro:2007sp} that rather low values of $\beta$ are needed to form a reasonably populated network of semilocal strings, and in this work we chose to perform the simulations for $\beta=0.01$, $\beta=0.04$ and $\beta=0.09$.The magnetic and scalar string cores for even lower  $\beta$ are too different in size and are difficult to simulate, since they are difficult to resolve and can overlap. Higher $\beta$ gives too scarce a network. 

We have performed simulations using two different scale factors ruling the expansion of the universe $t^\lambda$: radiation (t$^{1/2}$)  and matter (t$^{2/3}$). Since we are using the fat-string algorithm, this amounts to changing the damping term in each simulation accordingly. All in all, we have performed 12 simulations for each combination of the following parameters
\begin{itemize}
\item $\delta x=$ (0.5, 1)
\item $\beta=$ (0.01, 0.04, 0.09)
\item Cosmological era $=$ (radiation, matter)
\end{itemize}

\begin{figure*}
\begin{center}
\includegraphics[width=15cm]{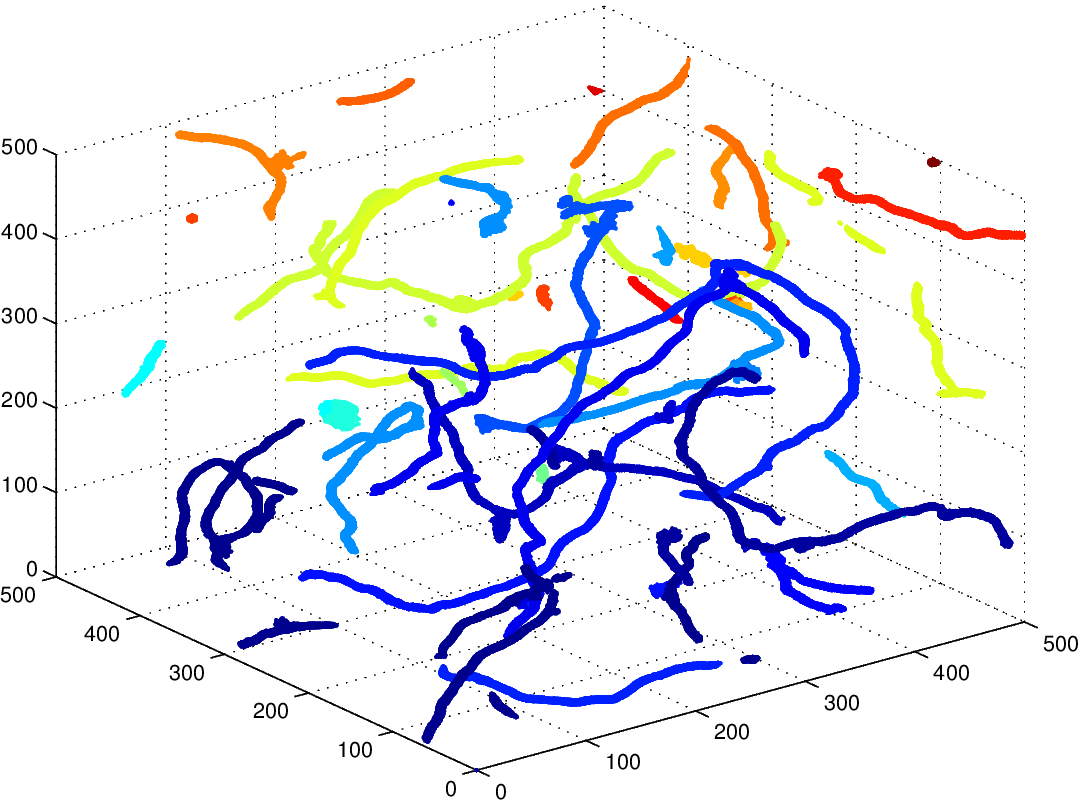}
\includegraphics[width=15cm]{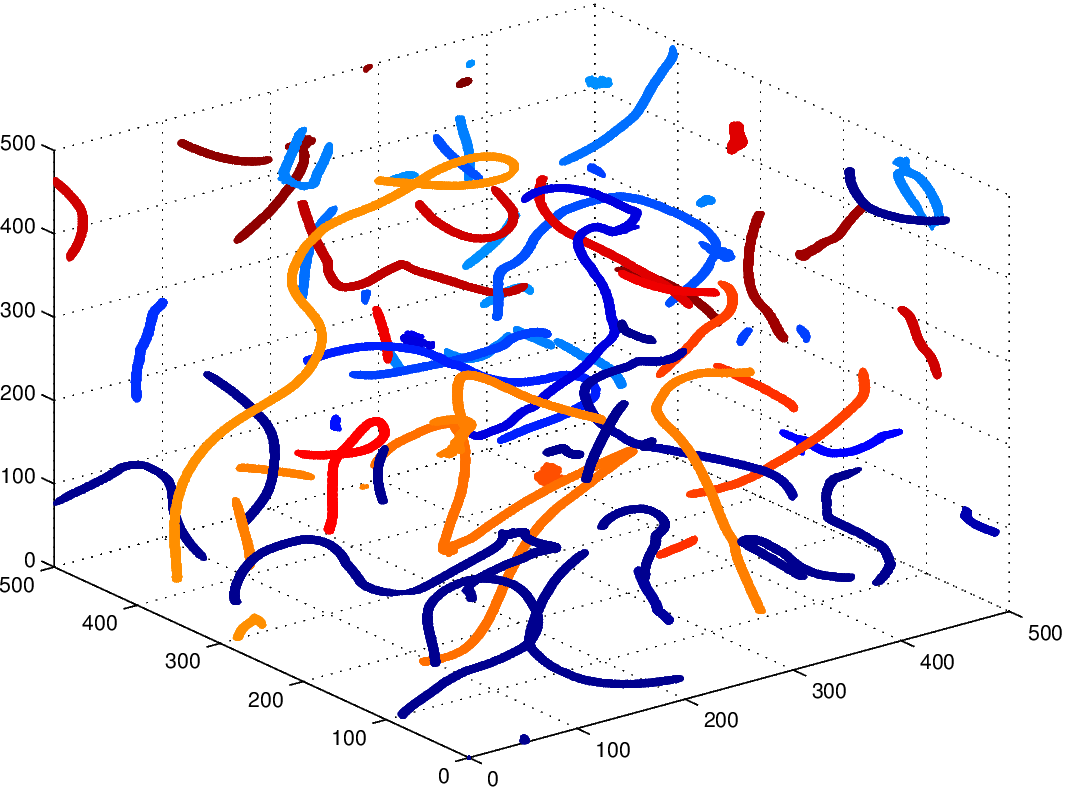}
\end{center}
\caption{Two snapshots of the simulations with $\beta=0.04$ in radiation (top) and matter (bottom). The top figure shows that strings in the radiation era are fuzzier, and many of the segments have energy lumps attached to them. The bottom figure shows segments that are in general smoother---an obvious consequence of the additional damping (there is less radiation in the box). Note that in either snapshot only part of the simulated box is depicted.}
\label{fuzzy}
\end{figure*}

There are several systematic errors that the reader should be aware of. On the one hand, there are numerical errors inherent to the simulation of the dynamics of the system. By these we mean errors arising because of the discretization of the equations (which will depend on the lattice spacing $\delta x$), errors coming from the fat-string algorithm and errors due to the limited dynamical range that can be obtained.

On the other hand, there are systematic errors in the identification and characterization of strings and monopoles from the simulation. The string lengths are obtained by dividing the string volume by the width of a static straight string, whereas our strings can be moving fast (and will therefore have Lorentz contraction), and have turns. In addition, the strings appear more fuzzy depending on the value of $\beta$ and the cosmology (meaning, amount of damping) we use. This fuzziness can sometimes be understood by considering that as strings move there is some radiation left behind, and if such lumps of radiation are {\it touching} the string, they are considered as string points by our algorithm. The end result is that the lower the value of $\beta$ and the smaller the damping term, the fuzzier the strings become, and also the bigger the energy blobs are.

An illustration of the effect of damping can be found in Fig.~\ref{fuzzy}, where we show snapshots of two simulations, one in radiation era and the other in matter era. In radiation, the strings appear to be more fuzzy, with some energy lumps attached to the strings; whereas in the matter era strings are noticeably smoother. Note that these cubes are only one part of the total simulation, which we have zoomed into to show the fuzziness more clearly; therefore the segments close to the boundaries would actually be continued in other parts of the box.

As for the effect of $\beta$, for lower $\beta$ the strings are expected to be more stable, as a result of the competition between gradient energy and potential energy \cite{Hindmarsh:1991jq}. Producing a blob without topology costs the same gradient and magnetic energy regardless of the value of $\beta$, but it does cost less potential energy for smaller $\beta$, thus again producing fuzzier strings.

As mentioned earlier, the number of monopoles is directly read from the number of segments. Some of the segments will in fact form a closed loop, so monopoles would be slightly overcounted by this procedure. Besides, even though we tried to factor out the energy blobs, some of them escaped our algorithm, and we are still counting those blobs as segments, and thus overcount monopoles again. Finally, the definition of segment is somewhat  arbitrary, since those segments that are not longer than $\alpha$ times the typical radius of a string are  discarded. Different choices for $\alpha$ can give different  number of segments.

Ways of quantifying some of these uncertainties will be briefly discussed in the following section, and in more detail in Papers II and III. For the moment, we provide one specific example, concerning the choice of the segment cutoff $\alpha$. We have analysed our data using $\alpha=1,3,5,8,10$, and we have found that the differences are bigger for smaller values of $\alpha$, but for $\alpha=5,8,10$ the results are more consistent. Also, this uncertainty decreases at later times, when there are fewer blobs and strings are longer. Tables~\ref{tables} and \ref{tablem}, which are described in the next section, show results for different $\alpha$.

Despite these numerical uncertainties, our methodology and sample size is sufficient to establish that the networks reach the expected scaling solution in all the cases studied. We discuss our results and compare them with our analytic models in the next section.

\section{Simulation results and comparison to analytic models}
\label{comp}

As described above, we have performed 12 simulations for each case of our set of parameters, and used results from the various sets of simulations  to obtain basic statistics about the properties of the networks. Each one of the 12 simulations in a given set has the same values for the parameters but a different initial random configuration, so that we can use them to obtain a purely statistical error. All in all, for each one of those simulations, and for specific values of the simulation time, we obtain the total string length ${\cal L}(t)$ and monopole number ${\cal N}(t)$ in the box. Both of these provide simple diagnostics for the large-scale evolution of the network, and specifically for the presence of scaling, as we will now discuss.

\begin{figure}
\begin{center}
\includegraphics[width=9cm]{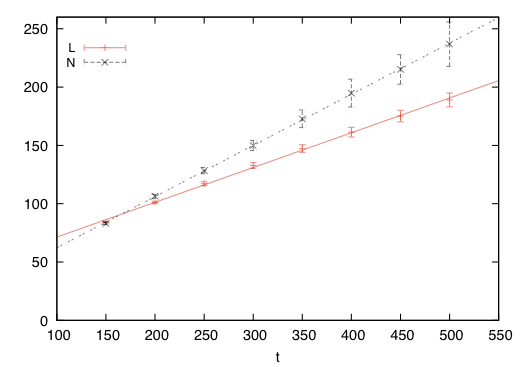}
\end{center}
\caption{Scaling plots for ${\cal L}$ and ${\cal N}$ for $\beta=0.09$, in the matter era, with $\delta x=1$. The error bars show statistical errors over the 12 simulations.}
\label{scale1}
\end{figure}

\begin{figure}
\begin{center}
\includegraphics[width=9cm]{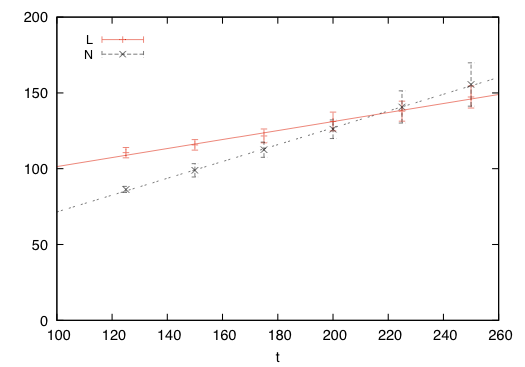}
\end{center}
\caption{Scaling plots for ${\cal L}$ and ${\cal N}$ for $\beta=0.01$, in the matter era, when $\delta x=0.5$. The error bars show statistical errors over the 12 simulations.}
\label{scale2}
\end{figure}

Figs.~\ref{scale1} and \ref{scale2} provide two examples of the evolution of these quantities, for the cases ($\beta=0.09$, $\delta x=1$) and ($\beta=0.01$, $\delta x=0.5$); both are matter era simulations. These are representative of all the sets of simulations we have performed. This analysis therefore shows that all the networks have reached the scaling solution by the corresponding final timesteps. The time needed for the different sets of networks to reach scaling is sligthly different, but this is to be expected given the different underlying conditions, such as the amount of damping in the simulation boxes. 

The obtained string lengths and number of monopoles can easily be translated into VOS-type lengthscales using Eq.~(\ref{linscal})
\be
\gamma_s\equiv\frac{L_s}{t}=\frac{1}{t}\sqrt{\frac{V}{\cal L}}\,,
\label{gs}
\ee
\be
\gamma_m\equiv\frac{L_m}{t}=\frac{1}{t}\left(\frac{V}{{\cal N}}\right)^{1/3}\,.
\label{gm}
\ee
It follows from our discussion in Section \ref{networks} that once a network reaches scaling both $\gamma_s$ and $\gamma_m$ should become constants. Note that while we do expect them to have comparable values of order (but slightly smaller than) unity, there is no expectation that they must be equal.

From our simulations, each of $\gamma_s$ and  $\gamma_m$ can be numerically calculated in various different ways. This turns out to be a simple but useful way of quantifying statistical and systematic uncertainties. We can average the values of $L_s(t)$ and $L_m(t)$  obtained in each simulation, and then calculate $\gamma_s$ and $\gamma_m$ using Eqns.~(\ref{gs}) and (\ref{gm}) on the averaged quantities; or, we can obtain one $\gamma_s$ and $\gamma_m$ for each simulation, and then average the $\gamma_s$ and $\gamma_m$ over all simulations. Moreover, if we are in (or approaching) scaling, the slopes of $L_s(t)$ or $L_m(t)$ evolution plots can also be used as numerical diagnostics for the corresponding $\gamma$. This is the prescription we actually use, by considering only the latter part of each set of simulations.

The result of both prescriptions is shown in Table~\ref{tables} for $\gamma_s$ and  Table~\ref{tablem} for $\gamma_m$. The values $\gamma_s\ (sim)$ show the values when the averaging has been done at the simulation level, and $\gamma_m\ (slope)$ when it is the $\gamma$'s which have been averaged. In all cases the errors quoted are statistical errors, which are smaller than the systematic error and so not directly indicative of the full uncertainty. Instead, they should be understood as lower bounds on the uncertainties in these simulations. 

Both Table~\ref{tables} and  Table~\ref{tablem} show comparisons of the simulations with $\delta x=1.0$ and $\delta x=0.5$. As mentioned before, the low value of $\delta x$ is a more accurate approximation to the continuous case, but lacks in dynamical range; whereas the higher value of $\delta x$ has a larger dynamical range though a poorer discretization.

The tables also show a comparison of the results obtained with two different values for the definition of segment; namely for $\alpha=3$ and $\alpha=8$. We investigated values of $\alpha=1,3,5,8,10$, and found that for the latter three the results are quite similar. The table shows that the magnitudes related to the string lengths do not change much with respect to the value of $\alpha$, whereas the monopole lenghtscale changes more. Not only the cases with a higher value of $\alpha$ are more similar to each other, but also the differences between $\delta x=0.5 $ and $\delta x=1.0$ are smaller for higher $\alpha$. Therefore, the systematics seem to be under better control for higher values of $\alpha$. 

\begin{table*}
\begin{center}
\begin{tabular}{c|c|c||c|c||c|c}
\hline
 $\beta$ & $\lambda$ & $\delta x$ & $\gamma_s (sim)(\alpha=3)$ &  $\gamma_s (slope)(\alpha=3)$ & $\gamma_s (sim)(\alpha=8)$  &$\gamma_s (slope)(\alpha=8)$ \\
\hline\hline
 0.01 & Rad & 0.5 & $ 0.27 \pm 0.02$& $ 0.27 \pm 0.04$ & $ 0.29 \pm 0.02$ & $ 0.29 \pm 0.04$ \\
 0.01 & Rad & 1.0 & $ 0.280\pm 0.003$ & $ 0.27 \pm 0.06$& $ 0.285 \pm 0.001$&  $ 0.29 \pm 0.01$ \\
\hline
 0.01 & Mat & 0.5 & $ 0.30 \pm 0.01$ & $0.30 \pm 0.05$ &$ 0.30 \pm 0.01$  &  $ 0.30 \pm 0.05$ \\
 0.01 & Mat & 1.0 & $ 0.292 \pm 0.002 $& $ 0.29 \pm 0.01$& $ 0.295 \pm 0.002$& $ 0.30 \pm 0.01$\\
\hline \hline
 0.04 & Rad & 0.5 & $ 0.301 \pm 0.005 $  & $ 0.30 \pm 0.04$ &$ 0.294 \pm 0.006$& $ 0.30 \pm 0.04$ \\
 0.04 & Rad & 1.0 & $ 0.283 \pm 0.004$& $ 0.28 \pm 0.01 $& $ 0.284 \pm 0.004$&  $ 0.28 \pm 0.01$ \\
\hline
 0.04 & Mat & 0.5 & $ 0.302 \pm 0.001 $ & $  0.30 \pm 0.03 $  &$ 0.301 \pm 0.001$&  $ 0.30 \pm 0.03$ \\
 0.04 & Mat & 1.0 & $ 0.291 \pm 0.005$ & $  0.29 \pm 0.01 $ & $ 0.291 \pm 0.005$  & $ 0.29 \pm 0.01$ \\
\hline\hline
 0.09 & Rad & 0.5 & $ 0.327 \pm 0.001$ & $ 0.33  \pm 0.05$ &$ 0.325 \pm 0.002$ & $ 0.33 \pm 0.05$ \\
 0.09 & Rad & 1.0 & $0.303 \pm 0.005$ & $ 0.30 \pm 0.01$ & $ 0.303 \pm 0.005$ & $ 0.30 \pm 0.01$ \\
\hline
 0.09 & Mat & 0.5 & $ 0.337\pm 0.006$ & $  0.33 \pm 0.07 $&  $ 0.336 \pm 0.006$ & $ 0.33 \pm 0.06$ \\
 0.09 & Mat & 1.0 & $ 0.307 \pm 0.006$ & $ 0.31 \pm 0.01$ & $ 0.306 \pm 0.006$ & $ 0.31 \pm 0.01$ \\
\hline
\hline
\end{tabular}
\caption{The measured values (with one-$\sigma$ statistical errors) of the string scaling parameter $\gamma_s$ for the various series of simulations described in the text.}
\label{tables}
\end{center}
\end{table*}

\begin{table*}
\begin{center}
\begin{tabular}{c|c|c||c|c||c|c}
\hline
  $\beta$ & $\lambda$ & $\delta x$ & $\gamma_m (sim)(\alpha=3)$ & $\gamma_m (slope)(\alpha=3)$ & $\gamma_m(sim)(\alpha=8)$ & $\gamma_m (slope)(\alpha=8)$ \\
\hline\hline
 0.01 & Rad & 0.5 & $ 0.549 \pm 0.007$ & $ 0.6 \pm 0.1$ &$ 0.586 \pm 0.005$&  $ 0.6 \pm 0.2$ \\
 0.01 & Rad & 1.0 & $ 0.34\pm 0.01$ & $ 0.34 \pm 0.02$ &$ 0.44 \pm 0.01$ &  $ 0.44 \pm 0.03$ \\
\hline
 0.01 & Mat & 0.5 & $ 0.544 \pm 0.007$ & $0.55 \pm 0.08$& $ 0.555 \pm 0.008$ & $ 0.56 \pm 0.08$ \\
 0.01 & Mat & 1.0 & $ 0.41 \pm 0.01 $ & $ 0.41 \pm 0.02$ &$ 0.47 \pm 0.01$ & $ 0.48 \pm 0.03$ \\
\hline\hline
 0.04 & Rad & 0.5 & $ 0.45 \pm 0.02 $ & $ 0.4 \pm 0.1$& $ 0.5 \pm 0.2$  &  $ 0.5 \pm 0.1$ \\
 0.04 & Rad & 1.0 & $ 0.359 \pm 0.009$ & $ 0.36 \pm 0.02 $ &$ 0.469 \pm 0.006$ & $ 0.46 \pm 0.02$ \\
\hline
 0.04 & Mat & 0.5 & $ 0.48 \pm 0.02 $ & $  0.47 \pm 0.09 $&  $ 0.49 \pm 0.01$ & $ 0.5 \pm 0.1$ \\
 0.04 & Mat & 1.0 & $ 0.424 \pm 0.006$ & $  0.43 \pm 0.02 $ &$ 0.466 \pm 0.004$ & $ 0.46 \pm 0.02$ \\
\hline\hline
 0.09 & Rad & 0.5 & $ 0.45 \pm 0.09$ & $ 0.45  \pm 0.09$& $ 0.46 \pm 0.02$  & $ 0.5 \pm 0.1$ \\
 0.09 & Rad & 1.0 & $0.397 \pm 0.007$ & $ 0.40 \pm 0.01$& $ 0.460 \pm 0.005$ & $ 0.46 \pm 0.02$ \\
\hline
 0.09 & Mat & 0.5 & $ 0.44\pm 0.01$ & $  0.45 \pm 0.06 $&  $ 0.45 \pm 0.02$ & $ 0.46 \pm 0.08$ \\
 0.09 & Mat & 1.0 & $ 0.419 \pm 0.004$ & $ 0.42 \pm 0.03$ &$ 0.442 \pm 0.003$ & $ 0.44 \pm 0.03$ \\
\hline
\hline
\end{tabular}
\caption{The measured values (with one-$\sigma$ statistical errors) of the monopole scaling parameter $\gamma_m$ for the various series of simulations described in the text.} 
\label{tablem}
\end{center}
\end{table*}

Given the way they were numerically determined, $L_s$ should be thought of as the typical inter-string distance (or perhaps the typical segment size), while $L_m$ is a characteristic inter-monopole distance. These are therefore not correlation lengths in the same strict sense as the term is used, for example, in Goto-Nambu string simulations. In particular, the fractal distribution of the semilocal networks (and more specifically the assumption of a Brownian network) is an issue that warrants further study. 

Bearing in mind the caveats we discussed, one should proceed with caution if trying to extract quantitative information from these scaling properties. (A further difficulty stems from the fact that we have as yet no accurate measurement of the defect velocities---this will be addressed in Paper III.) Nevertheless, it is encouraging that the overall behavior is in agreement with our understanding of the relevant underlying physical mechanisms. Specifically, we note that
\begin{itemize}
\item  For a given cosmology (damping term), $\gamma_s$ grows with $\beta$ and $\gamma_m$ gets smaller. This is to be expected since for lower $\beta$ we expect the system to behave more like an Abelian-Higgs network, which has longer strings and fewer segments (note that $\gamma_s$ and $\gamma_m$ are inversely proportional to ${\cal L}$ and ${\cal N}$, respectively). Analogous results have recently been found for cosmic strings \cite{Hiramatsu}.
\item For a given $\beta$, $\gamma_s$ is higher for higher damping terms, and $\gamma_m$ is lower. This is also to be expected since a lower damping term means that monopole velocities will be higher. Segments can therefore move faster to either grow and meet with other segments or collapse, giving a longer typical string length and smaller number of monopoles. One naturally expects that the additional length lost by segment collapse is more than compensated by that gained by the extra growth. (Note that increasing the string correlation length $L_s$ corresponds to decreasing the string density, and therefore the total length in string.)
\item One set of simulations (corresponding to radiation era, $\delta x=1$ and $\beta=0.01$) is an outlier, in the sense that it doesn't obviously follow the above trends. However, we note that this is the case where there is a smaller effective damping (and therefore more radiation) in the simulation box, and hence this is also the case that is most vulnerable to hidden systematics.
\end{itemize}

We should also point out that the scaling properties we have obtained for the string segments and monopoles are somewhat less sensitive to the value of $\beta$ than one might have expected. It is possible that this is a feature of the PRS algorithm, as has been recently discussed in  \cite{Hiramatsu}. Nevertheless, our results are consistent with an earlier set of semilocal simulations, discussed in \cite{Nunes}.

As in the case of the analysis in \cite{Nunes} a full direct calibration of the parameters of the analytic model for the evolution of the overall network cannot be done until we can numerically determine the velocities of the monopoles and segments---a task which we leave for paper III. Still, we can use the results of Table \ref{tablem} to provide a preliminary comparison with the model, and specifically with the scaling solution described by Eqs. (\ref{nscaling1}-\ref{nscaling2}). We will neglect the $\beta$ dependence, which as we saw is numerically found to be quite small when allowing for statistical and systematic uncertainties, and we will focus on the results for the $\alpha=8$ case for the reasons discussed above.

With these assumptions our free parameters are the analytic model parameters, $c_\star$ and $k_m$, as well as the monopole scaling velocities in the radiation and matter eras, which we will denote $v_{\rm rad}$ and $v_{\rm mat}$. Using our numerically determined values of $\gamma_m$, we find
\be
v_{\rm rad}\sim0.48 k_m\,
\ee
\be
v_{\rm mat}\sim0.20 k_m\,;
\ee
we have deliberately not included error bars in these numbers since we are unable to quantify possible systematic uncertainties in the $\gamma$'s. These values are consistent with the results of our earlier simulations \cite{Nunes}, where for a faster expansion rate ($\lambda=3/4$) we had found
\be
v_{\rm fast}\sim0.12 k_m\,.
\ee
As expected, faster expansion rates lead to smaller velocities. On the assumption that the analytic model is correct, we therefore infer that the ratio of the scaling monopole velocities in the matter and radiation eras should be
\be
\frac{v_{\rm mat}}{v_{\rm rad}}\sim0.4\,.
\ee

If one assumes a curvature parameter $k_m$ of order unity as in the case of Goto-Nambu strings \cite{VOS02}, our estimated velocities are comparable to (though possibly somewhat lower than) the ones typically encountered in other field theory defect simulations \cite{Moore,WallSim}. Thus, even though this comparison is somewhat simplistic, the results are at least encouraging. A full comparison (and thus a proper calibration of the analytic model) requires the numerical implementation of a reliable method to measure defect velocities in our simulations, which will be the subject of Paper III.

\section{Conclusions}

We took advantage of recent progress in computing facilities to carry out a more detailed numerical study of the evolution of semilocal string networks, the first results of which have been discussed above. These are based on the largest and most accurate field theory simulations of these objects to date, with sets of $1024^3$ simulations. Several of these sets have been simulated, thereby exploring a parameter space spanning different cosmological eras, values of the coupling $\beta$ and spatial resolutions, as well as thresholds for identification of the semilocal segments.

In the present work we have focused on the large-scale properties of these networks, our main result being a confirmation of earlier indications that linear scaling (analogous to the well-known one for cosmic strings) is the attractor solution for the entire parameter space of initial conditions that we have been able to reliably probe. A brief comparison of our numerical results with the predictions of a previously developed one-scale model for the overall evolution of these networks \cite{Nunes} is encouraging, though a proper comparison must be left for future work. We found the dependence of the scaling parameters on the coupling $\beta$ to be somewhat weaker than one may have naively anticipated. This may be a side-effect of our usage of the 'fat strings' algorithm \cite{Press:1989yh}, as recently discussed in a different context in \cite{Hiramatsu}.

As previously mentioned, the dynamics of these networks is more complex than that of plain Goto-Nambu strings, and therefore it cannot be fully described by a simple analytic model for the overall defect density. This must be complemented by a description of the evolution of the distribution of the individual semilocal segments. Indeed, the fact that the overall energy density of the network is scaling (which is, physically, what is being quantified by the evolution of $L_s$ or $L_m$) does not by itself ensure that the segment distribution is also scaling. In this sense one can say that a one-scale model is not sufficient to describe the full evolution of the network. To some extent this is analogous to the presence of small-scale structures on cosmic string networks, which can be characterized in Goto-Nambu simulations \cite{Fractal}.

The characterization of the semilocal segment population will be the subject of Paper II. Indeed, a new way of detecting segments should also be a good way to improve on the possible systematic uncertainties which have been discussed above. Segment identification is clearly the dominant contribution to these, and therefore this is one of the limiting factors preventing a more accurate calibration of the analytic model. The other current bottleneck is a reliable method of measurement of the defect velocities. Note that the two are to some extent related, since velocity measurements will in principle require the direct detection of the positions of the monopoles. Some possible methods to carry out these measurements will be presented and discussed in Paper III, leading to a deeper comparison between the analytic models and the numerical simulations, and thus to a proper calibration of the models themselves.

\begin{acknowledgments}
This work was done in the context of the project PTDC/FIS/111725/2009 from FCT, Portugal. The work of A.A. was supported by the Marie Curie grant FP7-PEOPLE-2010-IEF-274326 at the University of Nottingham, and by a Nottingham Research Fellowship. C.J.M. is supported by an FCT Research Professorship, contract reference IF/00064/2012, funded by FCT/MCTES (Portugal) and POPH/FSE (EC). A.L-E. and J.U. acknowledge financial support from the University of the Basque Country (EHUA 12/11), Basque Government (IT-559-10), the Spanish Ministry (FPA2009-10612, FPA2012-34456), and the Spanish Consolider-Ingenio 2010 Programme CPAN (CSD2007-00042). A.L-E. is also supported by the Basque Government grant BFI-2012-228.

Our numerical simulations were performed on the COSMOS Consortium supercomputer (within the DiRAC Facility, jointly funded by STFC and the Large Facilities Capital Fund of BIS-UK) and on the Milipeia cluster at the Laboratory for Advanced Computing at University of Coimbra.
\end{acknowledgments}

\bibliography{semilocal}
\end{document}